# Miniature Probe for Optomechanical Focus-adjustable Optical-resolution Photoacoustic Endoscopy

Zhendong Guo, Zhanhong Ye, Weihao Shao, Lili Jing, and Sung-Liang Chen

*Abstract*—Photoacoustic microscopy (PAM) is a promising imaging modality because it is able to reveal optical absorption contrast in high resolution on the order of a micrometer. It can be applied in an endoscopic approach by implementing PAM into a miniature probe, termed as photoacoustic endoscopy (PAE). Here we develop a miniature focus-adjustable PAE (FA-PAE) probe characterized by both high resolution (in micrometers) and large depth of focus (DOF) via a novel optomechanical design for focus adjustment. To realize high resolution and large DOF in a miniature probe, a 2-mm plano-convex lens is specially adopted, and the mechanical translation of a single-mode fiber is meticulously designed to allow the use of multi-focus image fusion (MIF) for extended DOF. Compared with existing PAE probes, our FA-PAE probe achieves high resolution of 3–5 μm within unprecedentedly large DOF of >3.2 mm, more than 27 times the DOF of the probe without performing focus adjustment for MIF. The superior performance is demonstrated by imaging both phantoms and animals including mice and zebrafishes *in vivo*. Our work opens new perspectives for PAE biomedical applications.

*Index Terms*—Depth of focus, endoscopy, focus adjustable, optical resolution, photoacoustic.

## I. INTRODUCTION

PHOTOACOUSTIC imaging (PAI) is a powerful imaging technique since it can provide non-invasive imaging with high resolution and high contrast. It has been wildly used in biomedical research [1]. PAI has three major implementations, including photoacoustic computed tomography (PACT) [2], photoacoustic microscopy (PAM) [3]–[9] and photoacoustic endoscopy (PAE) [10]–[27]. Among the three implementations, the miniature-probe-based PAE can be inserted into bodies to acquire the images of internal organs and their structural and functional information by the spectroscopic imaging capability of PAI [12]. Recently, PAE has demonstrated a number of applications, such as intravascular [14], [17], [18], gastrointestinal tract [11], [12], [16], [19], [20], [23], [25]–[27], and urogenital system [10], [13], [15], [24] imaging. In terms of how spatial resolution is determined, PAM (and PAE) can be further categorized into two types, namely acoustic-resolution PAM (AR-PAM) [3] and optical-resolution PAM (OR-PAM) [4]–[9] (and similarly, acoustic-resolution PAE (AR-PAE) [10]–[12], [15], [17], [18], [21] and optical-resolution PAE (OR-PAE) [13], [16], [19]–[27]). In AR-PAE, an unfocused or focused transducer is employed to provide spatial resolution, and the lateral resolution is limited to tens to hundreds of micrometers. By contrast, OR-PAE enables high lateral resolution up to several micrometers by using a focused laser beam at the expense of penetration depth. Thus, OR-PAE is highly promising for resolving fine features of tissue, such as single capillaries. However, depth of focus (DOF) is reduced drastically as the spot size of the focused laser beam decreases (i.e., for high lateral resolution). As a result, when the OR-PAE probe using a tightly focused laser beam is employed to image internal organs, only the tissue within the very limited DOF can enjoy high microscale resolution, while that outside the DOF that may also contain fine structures cannot be well visualized. Further, in clinical applications, the shape of the inner surface of internal organs is typically irregular. In this regard, the image quality would be highly hampered due to limited DOF.

Currently, several OR-PAE probes with microscale resolution (<10 μm) have been demonstrated [16], [21], [22], [24]. These probes suffer from very limited DOF and may restrict clinical endoscopic imaging applications. Efforts have also been made to extend DOF or enable focus adjustment for PAE with resolution of tens of micrometers. An auto-focusing OR-PAE probe was fabricated to solve the deterioration of lateral resolution in the out-of-focus region for usually irregular gastrointestinal tract imaging [23]. However, the resolution is limited to 49 μm, and the probe diameter of 9 mm is large mainly due to the use of a 6-mm liquid lens. Another OR-PAE probe has achieved large DOF of ~8.6 mm in air by producing Bessel beams using an elongated focus lens, yet it remains to have low resolution of ~40 μm and large probe diameter of 8 mm [25]. Very recently, by using scanning-domain synthesis of optical beams, PAE with high resolution of 11 μm with DOF of 1.88 mm has been demonstrated [27], still the microscale resolution, typically achieved in OR-PAM, is not realized. Moreover, the probe diameter of 5 mm is relatively large. Large probe size may impede some medical applications such as intravascular imaging. Several methods to extend DOF of OR-PAM have been proposed [5]–[9]. A motorized stage was

Manuscript received XXX. *(Corresponding author: S. L. Chen.)*
This work was supported by National Natural Science Foundation of China (NSFC) under Grant 61775134.
Z. Guo, Z. Ye, and S. L. Chen are with University of Michigan-Shanghai Jiao Tong University Joint Institute, Shanghai Jiao Tong University, Shanghai 200240, China (e-mail: sungliang.chen@sjtu.edu.cn).
W. Shao and L. Jing are with School of Pharmacy, Shanghai Jiao Tong University, Shanghai 200240, China.

used to scan the imaging head along the axial direction (i.e., depth scanning) [5], [7], yet the approach cannot be used in PAE due to limited space in internal organs. Electrically tunable lenses were also employed to adjust the focus [6], [8], [9]. However, the imaging head integrating such a tunable lens is bulky and cannot be adopted in miniature PAE probes. Hence, a new design to simultaneously achieve microscale resolution, large DOF, and miniature probe size (<3 mm) remains a challenge and is worth investigating.

Here we present novel a focus-adjustable PAE (FA-PAE) probe. The focus can be adjusted by controlling the distance between a single-mode optical fiber (SMF) and a 2-mm plano-convex lens. Then, by fusing photoacoustic A-line signals (or images) at different focal planes, DOF is equivalently extended. This approach is also termed as multi-focus image fusion (MIF). For PAE imaging, the probe achieves unprecedented performance in terms of both high resolution of 3−5 μm and large DOF of >3.2 mm, more than 27 times the DOF of the probe without performing focus adjustment for MIF. The outer diameter is 2.9 mm, which facilitates clinical PAE applications. Rotary scanning for cross-sectional imaging is performed to show the feasibility of using the probe in endoscopy settings. Further, *in vivo* imaging of mice and zebrafishes is conducted to demonstrate the superior imaging performance of the probe. Compared with previous PAE probes, our FA-PAE probe offers miniature size and high microscale resolution over large DOF, which are highly desired for clinical PAE applications. It is worth mentioning that the novel optomechanical design of focus adjustment is immune against electromagnetic interference, which is critical for intravascular imaging.

## II. METHODS

In OR-PAE (or OR-PAM), the optical diffraction-limited lateral resolution is expressed as:

$$\text{Resolution} = 0.51 \frac{\lambda}{\text{ENA}}, \quad (1)$$

where $\lambda$ denotes the laser wavelength, and ENA denotes the effective numerical aperture (NA) of the focused laser beam. ENA is determined by the expression:

$$\text{ENA} = n \frac{D}{2f}, \quad (2)$$

where $n$ is the refractive index of the medium where the lens is working, $D$ is the laser beam size on the lens, and $f$ is the focal length. In our previous work [22], it was demonstrated that the $f$ and ENA (and thus lateral resolution) are adjustable by changing the distance, $d$, between an SMF and a focusing lens, as shown in Fig. 1. Therefore, it is possible to adjust $f$ by changing $d$ in an OR-PAE probe. On the other hand, since ENA also varies when changing $d$, numerical simulation is needed to better understand the ENA (and thus lateral resolution) and $f$ as a function of $d$. Specifically, there is a trade-off between lateral resolution and $f$ when changing $d$. Zemax was used to perform the simulation. The parameters used in our probe (described below) were chosen in the simulation.

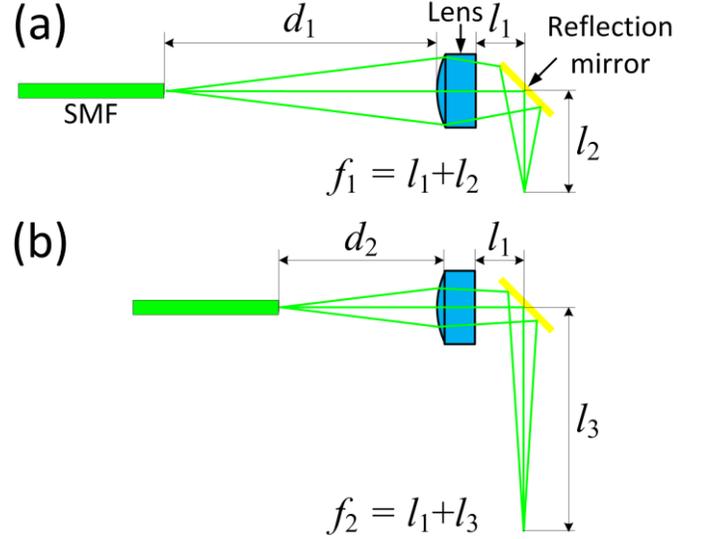

Fig. 1. Illustration of adjustable $f$ by changing $d$. When $d$ is reduced from (a) to (b), i.e., $d_2 < d_1$, $f$ is increased, i.e., $f_2 > f_1$. Note that the ENA is smaller, and thus, lateral resolution is lower in (b).

Figure 2(a) shows the schematic of the FA-PAE probe in cross-sectional and in three-dimensional (3D) views. An SMF (S405XP, Nufern) was firstly fixed in a long plastic tube (PT1; inner diameter (ID): 0.3 mm; outer diameter (OD): 1 mm). Then, PT1 was inserted into another long plastic tube (PT2; ID: 1.1 mm; OD: 2 mm) without fixation. At the proximal end, a glass tube (GT1; ID: 1.1 mm; OD: 1.8 mm) was slid in PT1 and fixed with it, and another glass tube (GT2; ID: 1.1 mm; OD: 2 mm) was slid in PT1 without fixation. A metal tube (MT1; ID: 2.1 mm; OD: 2.5 mm) that has a ~270° side window opened at a section of MT1, as indicated in Fig. 2(a), was further slid in the proximal end and fixed with GT2 and PT2. On the other hand, at the distal end, another glass tube (GT3; ID: 0.3 mm; OD: 1.7 mm) was slid in the SMF and fixed with it. That is, the SMF, PT1, GT1, and GT3 were fixed together, which is called the moving unit. Another metal tube (MT2; ID: 1.8 mm; OD: 2 mm) was further slid in GT3 and fixed with PT2. Note that MT2 and GT3 were not fixed. A 2-mm diameter plano-convex lens (43-397, Edmund) was then fixed at the distal end of MT2. Another metal tube (MT3; ID: 2.1 mm; OD: 2.7 mm) that has a 45° end face and a ~180° side window opened at the distal end of MT3 was slid in MT2. A home-made gold-coated thin film (GCF) (48-1F-OC, CS Hyde) was attached at the 45° end face of MT3. Finally, an ultrasonic transducer (AT23730, Blatek) with miniature dimensions (0.6 mm × 0.5 mm × 0.2 mm) was attached at the distal end along MT3. The transducer has central frequency of ~40 MHz with bandwidth of ~60% and was used for acoustic detection. The GCF is for light reflection (for side-view imaging) and sound transmission. The 180° side window is to allow light and sound transmission. To realize focus adjustment, GT1 was connected to a one-dimensional (1D) motorized stage (shown later) for linear motion of the moving unit and thus the SMF. Other than the moving unit, other components were kept stationary during the process of focus adjustment. As a result, $d$ and thus $f$ can be changed to achieve focus adjustment. Note that GT3 was used to ensure good coaxial alignment of the SMF with the lens during the process of focus adjustment. PT1 and PT2 were used for flexible bending of the probe, which facilitates clinical PAE applications. For the fixation of different components mentioned above, UV epoxy was used. Figures 2(b)−2(e) show the pictures of the FA-PAE probe. As can be seen in Fig. 2(e), the probe diameter is 2.9 mm.

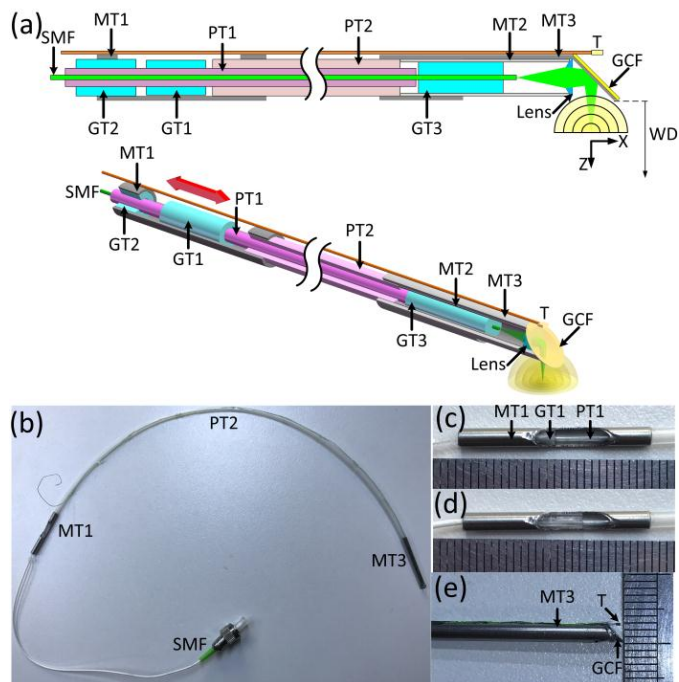

Fig. 2. (a) Schematic of the FA-PAE probe in cross-sectional and 3D views. (b) Picture of the whole probe to show the long PT2 for flexible bending. (c,d) Picture of the proximal end of the probe to show linear motion of the moving unit. $d$ is reduced from (c) to (d), i.e., $f$ is increased from (c) to (d). (e) Picture of the distal end of the probe to show the miniature size. T, transducer; WD, working distance.

The schematic of the imaging system for the FA-PAE probe is shown in Fig. 3. A Q-switched diode-pumped solid-state laser (SPOT-10-200-532, Elforlight, UK) was employed to provide pulsed laser (pulse duration: <2 ns, pulse repetition frequency: 1 kHz, wavelength: 532 nm) for photoacoustic excitation. A variable neutral density filter was used to adjust the laser energy, and an iris to obtain a more circular beam shape. Then, a beamsplitter was used to split the laser, where a small portion of the laser was fed into a photodetector (DET10A, Thorlabs) for triggering, and the majority of the laser was spatially filtered and coupled into the SMF of the FA-PAE probe via a fiber coupler (F-915T, Newport). A 1D motorized stage (M-404, Physik Instrumente [PI], Karlsruhe, Germany) was used to realize linear motion of the SMF, which controls $d$ and thus $f$ for focus adjustment. Photoacoustic signals were detected by the transducer. Then, the photoacoustic signals were amplified by a preamplifier (ZFL-500LN-BNC+, Mini-Circuits) and an ultrasonic pulser/receiver (5073PR, Olympus) successively. The amplified signals were sampled by a high-speed digitizer (CSE1422, GaGe) with sampling rate of 200 MS/s and 14-bit resolution. The data were saved to a personal computer for further signal processing and image display. The computer was also used to synchronize the pulsed laser, the probe scanning, and the data acquisition. To demonstrate the imaging capability of the probe, both linear and rotary scanning schemes were implemented. For linear scanning, the probe was mounted on a two-dimensional (2D) motorized stage (M-404, Physik Instrumente [PI]). This is more convenient to demonstrate the superior imaging performance of the probe. For rotary scanning, the sample was rotated by using a step motor (not shown in Fig. 3), while the probe was linearly scanned along the axial direction. This is to demonstrate the feasibility of our probe in acquiring cross-sectional images for clinical PAE applications in future. The sample was mounted on a 3D stage to facilitate the alignment in experiment.

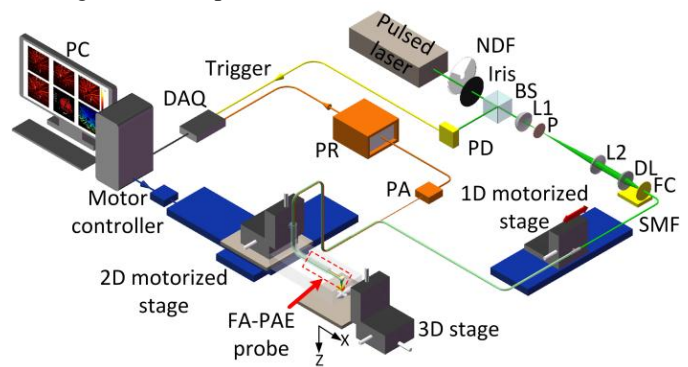

Fig. 3. Schematic of the imaging system for the FA-PAE probe. NDF, neutral-density filter; BS, beamsplitter; PD, photodetector; L1, lens 1; L2, lens 2; P, pinhole; DL, doublet lens; FC, fiber coupler; PA, preamplifier; PR, pulser/receiver; PC, personal computer.

### III. RESULTS

#### A. Resolution and DOF

As mentioned previously, Zemax was used to simulate the changes of $f$, ENA, and lateral resolution as a function of $d$. Firstly, the $f$ vs. $d$ curve can be plotted by Zemax simulation result directly, as shown in Fig. 4(a). Then, the ENA vs. $d$ curve can be obtained by using (2), as shown in Fig. 4(b). Finally, the lateral resolution vs. $d$ curve can be further calculated by using (1), as shown in Fig. 4(c). Besides, we measured the $f$ and ENA by changing $d$ in experiment, and plotted the result in Figs. 4(a) and 4(b). As can be seen, both $f$ and ENA show excellent agreement between simulation and experiment. To measure lateral resolution at different $d$, we conducted experiment by imaging the sharp edge of a razor blade immersed in water at different $d$. The scanning step size was 0.5 µm. A 1D photoacoustic amplitude profile was obtained and fitted by a sigmoidal-shaped curve as the edge spread function (ESF) of the profile. A line spread function (LSF) can be calculated by taking the spatial derivative of the ESF [22]. The resolution is determined by checking the FWHM of the LSF. Figure 4(d) shows the representative result when $d$ = 5.5 mm, and the lateral resolution of ~3.0 µm was measured. All measured lateral resolution at different $d$ is also plotted in Fig. 4(c). Figure 4(c) shows slightly worse experimental result compared with the simulation one. This is very likely because the laser beam was distorted after reflected by the non-perfectly flat surface of the GCF. Based on the results in Figs. 4(a) and 4(c), lateral resolution vs. $f$ curve can be plotted, as shown in Fig. 4(e), which indicates that the FA-PAE probe can achieve high resolution of 3−5 μm over large DOF of >3.2 mm ($f$ = ~3.0–6.2 mm) based on the MIF approach. On the other hand, the intrinsic DOF based on a single focus (SF) was experimentally measured at a fixed $f$ (and $d$), i.e., without focus adjustment. Specifically, at a certain $f$, lateral resolutions at different depths were measured by the method the same as Fig. 4(d). Then, the intrinsic DOF can be determined. As shown in Fig. 4(e), the intrinsic DOF was estimated to be 38 μm and 118 μm at two representative f of ~3.0 mm and ~6.2 mm. As a comparison, the MIF-based DOF is significantly improved by more than 27 times compared with the SF-based DOF for SF-based lateral resolution of <5 μm. Note that the laser path distance from the lens to the outer boundary of the probe was 2.4 mm (i.e., the sum of the distance from the lens to the GCF and that from the GCF to the boundary of the probe). Thus, the working distance (WD, also see Fig. 2(a)) of the probe was ~0.6−3.8 mm. That is, WD = $f$ − 2.4. The axial resolution of the probe was mainly determined by the transducer's acoustic bandwidth. A 6-μm carbon fiber was imaged, and the photoacoustic A-line signal is shown in Fig. 4(f). Hilbert transform (envelope detection) was applied to the A-line signal, and then, the envelope was fitted by a Gaussian curve. Finally, the axial resolution was determined to be 45 μm by checking the FWHM of the Gaussian curve.

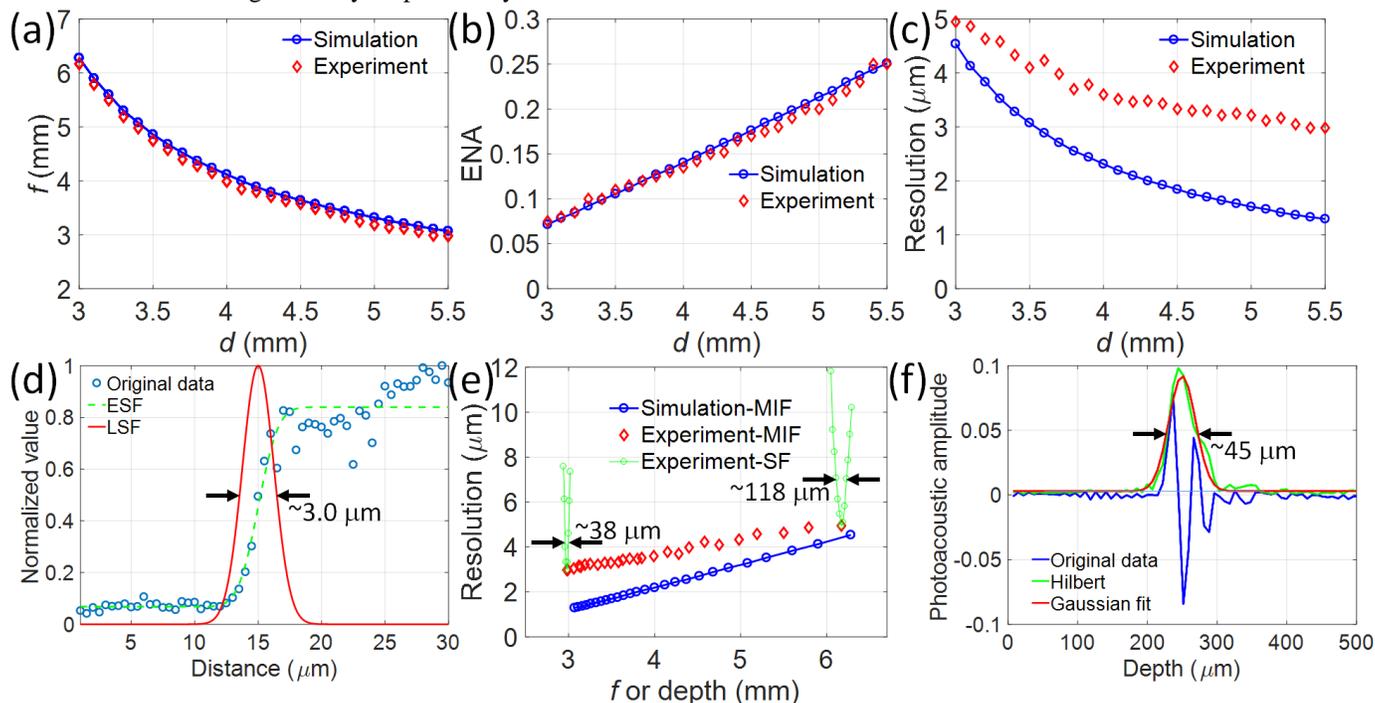

Fig. 4. $f$ (a), ENA (b), and lateral resolution (c) as a function of $d$. (d) Measurement of lateral resolution at $d$ = 5.5 mm. (e) Lateral resolution and DOF for MIF-based (simulation and experiment) and SF-based (experiment) cases. (f) Measurement of axial resolution.

*B. Phantom Imaging*

A phantom consisted of several 6 μm carbon fibers distributed in different depth was prepared and imaged by the FA-PAE probe. The photoacoustic images were obtained at three focal planes of $f$ = ~3.3, ~4.5, and ~5.9 mm (i.e., WD = ~0.9, ~2.1, and ~3.5 mm) by focus adjustment. Figure 5(a) shows the 3D rendering MIF image, and Fig. 5(b) shows the SF counterpart at $f$ = ~4.5 mm. As expected, all carbon fibers can be clearly resolved in Fig. 5(a), whereas only partial carbon fibers are displayed in Fig. 5(b). Specifically, the carbon fibers at the deep region (i.e., the bottom part in Figs. 5(a) and 5(b)) disappear in Fig. 5(b). Besides, although the carbon fiber at the shallow region (i.e., the top part) is observed in both Figs. 5(a) and 5(b), it is much blurred in Fig. 5(b) due to the limited DOF of the SF image. For further comparison, photoacoustic XY maximum amplitude projection (MAP) images of the MIF and SF images are shown in Figs. 5(c) and 5(d), respectively. Figures 5(e) and 5(f) show the zoom in images of the dashed boxes in Figs. 5(c) and 5(d), respectively. Figure 5(g) shows the comparison of the 1D profiles along the dashed lines in Figs. 5(e) and 5(f). From Figs. 5(e)-5(g), it can be clearly observed that the MIF image preserves both high resolution and SNR.

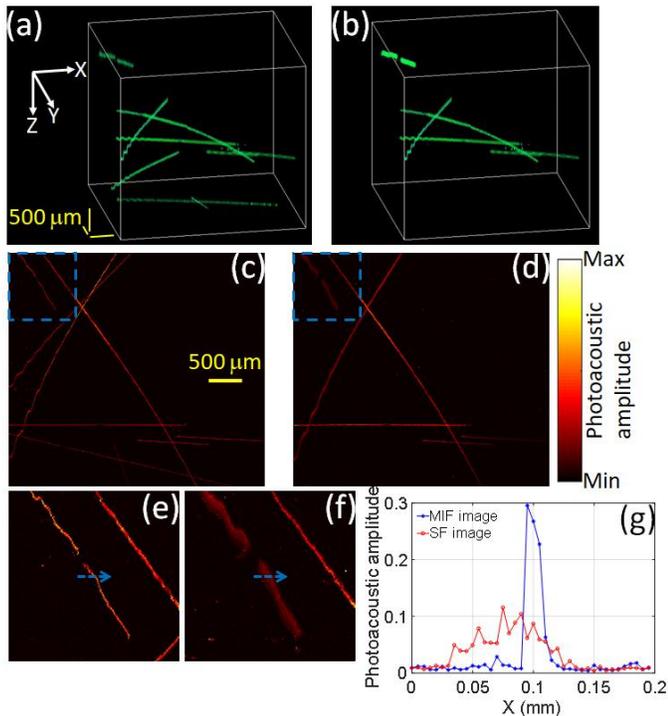

Fig. 5. Imaging of spatially distributed carbon fibers. 3D rendering MIF (a) and SF (b) images. Photoacoustic XY MAP images of the MIF (c) and SF (d) images. (e,f) The zoom in images of the dashed boxes in (c) and (d), respectively. (g) Comparison of the 1D profiles along the dashed lines in (e) and (f). The Z range in (a) and (b) corresponds to WD of 0.9–3.6 mm. (a) and (b) share the same scale bar in (a). (c) and (d) share the same scale bar in (c).

Another phantom of a leaf skeleton dyed with black ink was imaged to show the imaging capability of large DOF by the FA-PAE probe. The dyed leaf phantom was covered by UV epoxy to prevent the ink from leaking out of the phantom. To showcase the large DOF of the probe, the phantom was obliquely placed with the left part shallower and the right part deeper. Figure 6(a) shows the picture of the sample, where the red box region was imaged. Figure 6(b) shows the photoacoustic XY MAP image of the MIF image after fusing the photoacoustic images at multiple focal planes at f in the range of ~3.1–5.9 mm. On the other hand, Fig. 6(c) shows the photoacoustic XY MAP image of the SF image at $f$ = ~4.6 mm. At a first glance, the patterns in the right part (corresponding to the deep region) can be better revealed by the MIF image. Figure 6(d) shows the depth-encoded image of Fig. 6(b). Note that in Fig. 6(d), "Z (mm)" in the color bar represents the distance from the focal plane of the SF image, and is the same for all depth-encoded images throughout this paper. The large imaging depth range of more than 3.3 mm is identified in Fig. 6(b). We further check the zoom in images of Figs. 6(b) and 6(c), as shown in Figs. 6(e)−6(g), which have the majority of the patterns around Z = −1.4 mm, 0 mm, and 1.4 mm, respectively. As expected, the MIF image shows high image quality at all the three layers, while the SF image only preserves the high quality at the layer of Z = 0 mm.

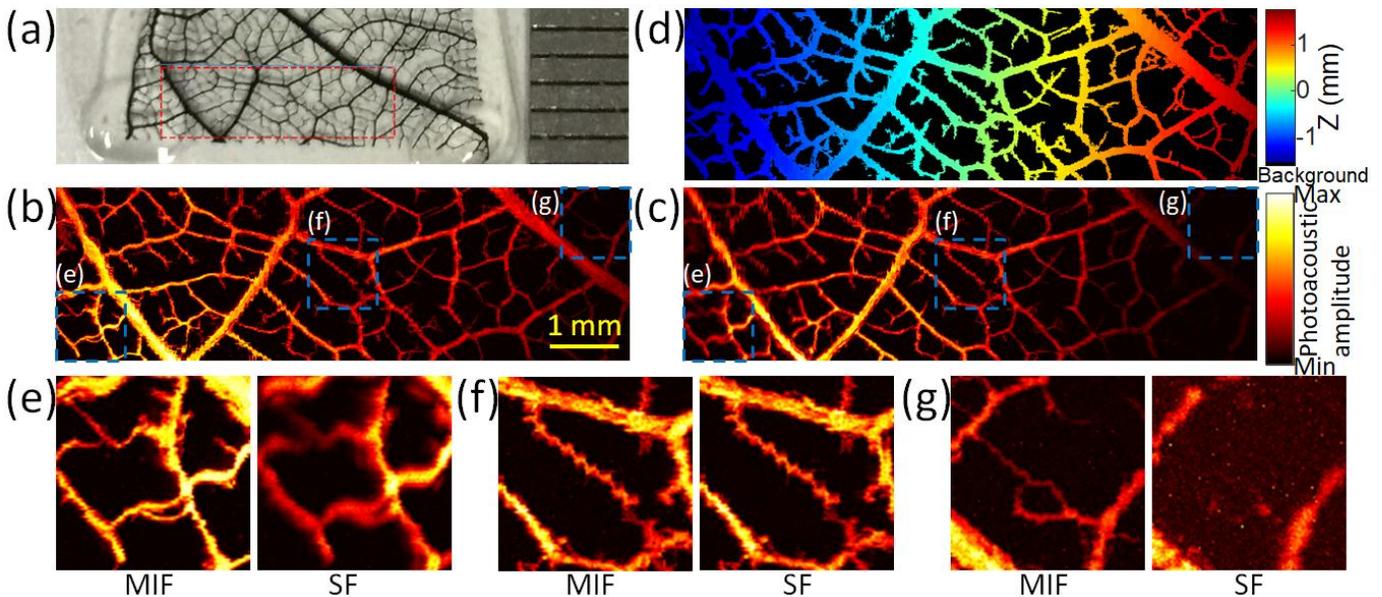

Fig. 6. (a) Photograph of the leaf phantom. Photoacoustic XY MAP images of the MIF (b) and SF (c) images. (d) The depth-encoded image of (b). Z represents the distance from the focal plane of the SF image ($f$ = 4.6 mm in this case). (e,f,g) Comparison of the zoom in images of the dashed boxes in (b) and (c). (e) for Z = −1.4 mm; (f) for Z = 0 mm; (g) for Z = 1.4 mm. (b)–(d) share the same scale bar in (b).

As mentioned previously, imaging based on rotary scanning was also conducted to show the feasibility to acquire cross-sectional images. Two sheets of dyed leafs were imaged. As shown in Fig. 7(a), the two leafs were placed around the probe with one at ~0° and the other at ~90° along the azimuthal axis. Note that the two sheets of leafs are relatively flat. During image acquisition, the phantom was rotated with angular step size of 0.225°, and the probe was moved along the Z axis. That is, 2D scanning was used for 3D imaging. Photoacoustic images at multiple focal planes at f of ~2.8−5.5 mm were acquired. Figures 7(b) and 7(c) show 3D rendering images and photoacoustic YZ MAP images, respectively, of the MIF and SF images (at $f$ = ~3.9 mm). Obviously, the MIF image delineates more leaf skeletons. Figure 7(d) shows the 2D XY slices at around the center of Z in Fig. 7(b). The 1D profiles along the dashed lines in Fig. 7(d) are plotted in Fig. 7(e). As can be seen in Fig. 7(e), three peaks can be easily distinguished in the MIF image by virtue of large DOF, while only two peaks are observed in the SF image. Moreover, the imaged size of the SF image is apparently larger (i.e., blurred). Note that for the case of rotary scanning, the XYZ mentioned above refers to the coordinates in Figs. 7(a) and 7(b). Otherwise, the coordinates in Figs. 2 and 3 are referred to.

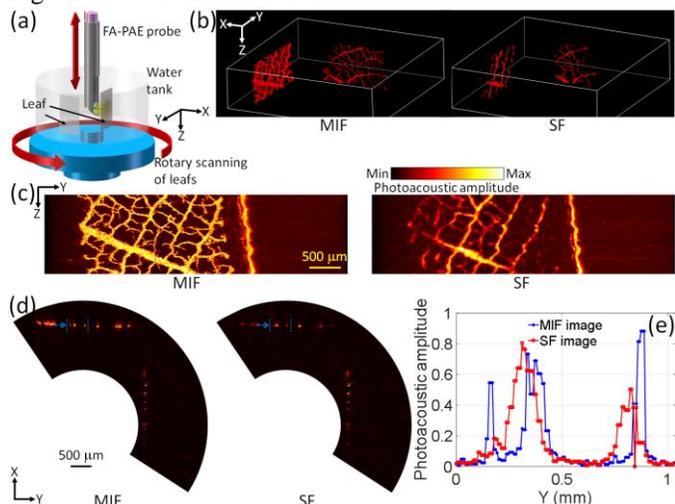

Fig. 7. (a) Schematic of the imaging based on rotary scanning. 3D rendering images (b) and photoacoustic YZ MAP images (c) of the MIF and SF images. 3D videos of the MIF (Video 1) and SF (Video 2) images are available. (d) 2D XY slices at around the center of Z in (b). (e) Comparison of the 1D profiles along the dashed lines in (d).

### C. In Vivo Imaging

To verify the *in vivo* imaging capability of the FA-PAE probe, we acquired the images of the mouse eye and ear. Before experiment, the mouse was anesthetized by injection of pentobarbital of 60 mg/kg and then fixed on a home-made animal platform. During experiment, all experimental animal procedures were carried out in conformity with the laboratory animal protocol approved by Laboratory Animal Care Committee of Shanghai Jiao Tong University.

The mouse eye is a spherical object by nature. When imaging the mouse eye, it is challenging to use conventional SF-based PAE to achieve high resolution over the whole pupil because of the spherical surface of the eyeball. By using the FA-PAE probe, the above issue can be elegantly addressed thanks to the extended DOF. Imaging of the mouse eye is a representative application of the probe. Figure 8(a) shows photoacoustic XY MAP images of the MIF and SF images. The MIF image was obtained by fusing photoacoustic images at focal planes of $f = $ ~3.1–3.7 mm, while the SF image was acquired at $f = $ ~3.7 mm. Figure 8(b) is the depth-encoded image of the MIF image in Fig. 8(a). As can be seen in Fig. 8(b), the top part is shallower, while the bottom deeper. Besides, the imaging depth range is ~0.9 mm. Figures 8(c) and 8(d) plot two representative zoom in images of the dashed boxes in Fig. 8(a). Figures 8(c) and 8(d) have the majority of the patterns around $Z = -0.6$ mm and $-0.5$ mm, respectively. As expected, blood vessels are clearly revealed in the MIF images, whereas those are much blurred in the SF images.

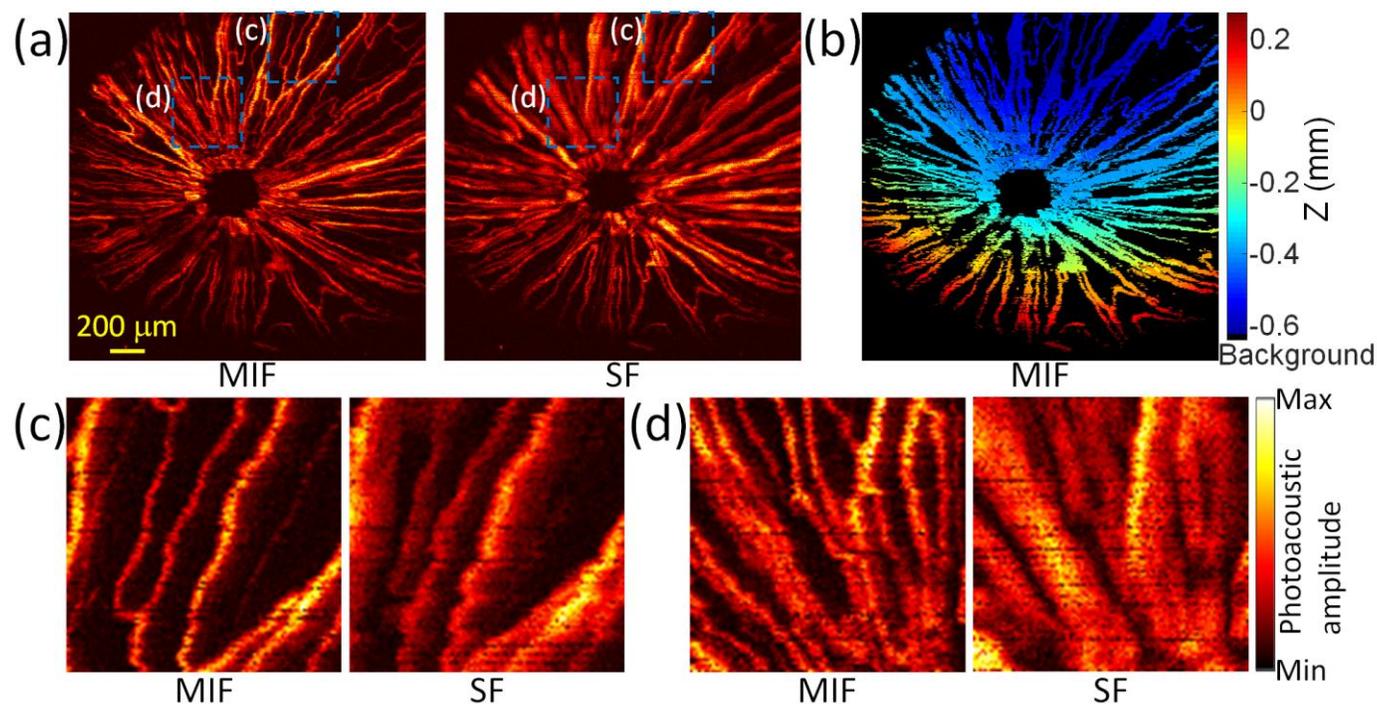

Fig. 8. (a) Photoacoustic XY MAP images of the MIF and SF images. (b) The depth-encoded image of the MIF image in (a). 3D videos of the MIF (Video 3) and SF (Video 4) images are available. Z represents the distance from the focal plane of the SF image ($f = 3.7$ mm in this case). (c,d) Comparison of the zoom in images of the dashed boxes in (a). (c) for $Z = -0.6$ mm; (d) for $Z = -0.5$ mm. (a) and (b) share the same scale bar in (a).

The mouse ear was also imaged by the FA-PAE probe. Figure 9(a) shows the picture of the mouse ear, where the dashed box indicates the imaging region. The ear was obliquely placed to make the left part shallower and the right part deeper. Figure 9(b) shows photoacoustic XY MAP images of the MIF ($f$ = ~3.3–5.5 mm) and SF ($f$ = ~4.5 mm) images. Apparently, the MIF image displays great image quality in high resolution over the whole imaging region by virtue of large DOF. In contrast, the SF image can only resolve fine structures in a limited region, specifically around the center part of the SF image in Fig. 9(b), and otherwise, blood vessels are highly blurred or even missing. Figure 9(c) presents the depth-encoded images of the MIF and SF images in Fig. 9(b). Similarly, the MIF image has high resolution over large depth range of ~2.6 mm. Figures 9(d) and 9(e) show the zoom in images of the dashed boxes in Fig. 9(b), where Figs. 9(d) and 9(e) have the majority of the patterns around Z = −1.1 mm and −0.4 mm, respectively. Further, the 1D profiles along the dashed lines are also checked (Figs. 9(d) and 9(e)). As can be seen, blood vessels are blurred with low SNRs (Fig. 9(e)) and even missing (Fig. 9(d)) in the SF image as Fig. 9(d) is further out of focus, while they are perfectly presented in the MIF image. The results manifest that the FA-PAE probe capable of producing MIF images can effectively solve the issues encountered by using conventional SF-based PAE probes. Moreover, the high resolution of our probe is evidenced by the imaged single capillaries and red blood cells (indicated by the white dashed box in Fig. 9(d)) in the MIF image.

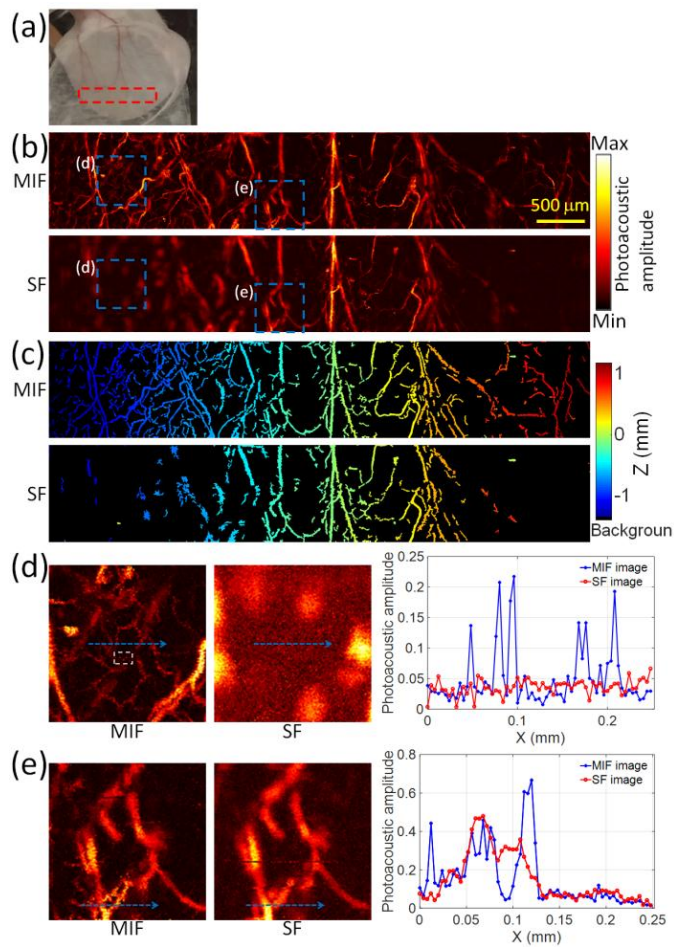

Fig. 9. (a) Picture of mouse ear. (b) Photoacoustic XY MAP images of the MIF and SF images. (c) The depth-encoded images of (b). Z represents the distance from the focal plane of the SF image ($f$ = 4.5 mm in this case). (d,e) Comparison of the zoom in images of the dashed boxes in (b); comparison of the 1D profiles along the dashed lines. (d) for Z = −1.1 mm; (e) for Z = −0.4 mm. Single red blood cells are indicated by the white dashed box in the MIF image in (d). (b) and (c) share the same scale bar in (b).

We further employed the FA-PAE probe to noninvasively image zebrafish larvae *in vivo*. A 30 dpf AB zebrafish with ~7 mm body length was anesthetized using 25x tricaine (4 mg/ml). Then, the zebrafish was carefully immobilized on a glass slide with low melt agarose (1.2 %) to keep it alive. The glass slide with a tilt angle of ~30° was mounted on the sample stage, which allows the demonstration of large imaging depth range. All experimental animal procedures were carried out in conformity with the laboratory animal protocol approved by Laboratory Animal Care Committee of Shanghai Jiao Tong University.

Whole-body imaging of living zebrafish larvae was acquired. Figure 10(a) depicts photoacoustic XY MAP images of the MIF ($f$ = ~3.4–5.7 mm) and SF ($f$ = ~4.6 mm) images. The depth-encoded images are also plotted in Fig. 10(b). As shown in Figs. 10(a) and 10(b), the MIF images render superior image quality because of high resolution over large DOF. The imaging depth range of the MIF image with high quality is ~2.7 mm. Two zoom in images (the dashed boxes in Fig. 10(a)) are checked at the regions with the majority of the patterns around Z = −1.1 mm and 1.1 mm, as shown in Figs. 10(c) and 10(d), respectively. Compared with the SF image, the MIF image is able to resolve fine structures, and the fish's mouth, eye, and tail can be clearly identified. The results also suggest that our probe holds promise to study other embryos and larvae *in vivo*.

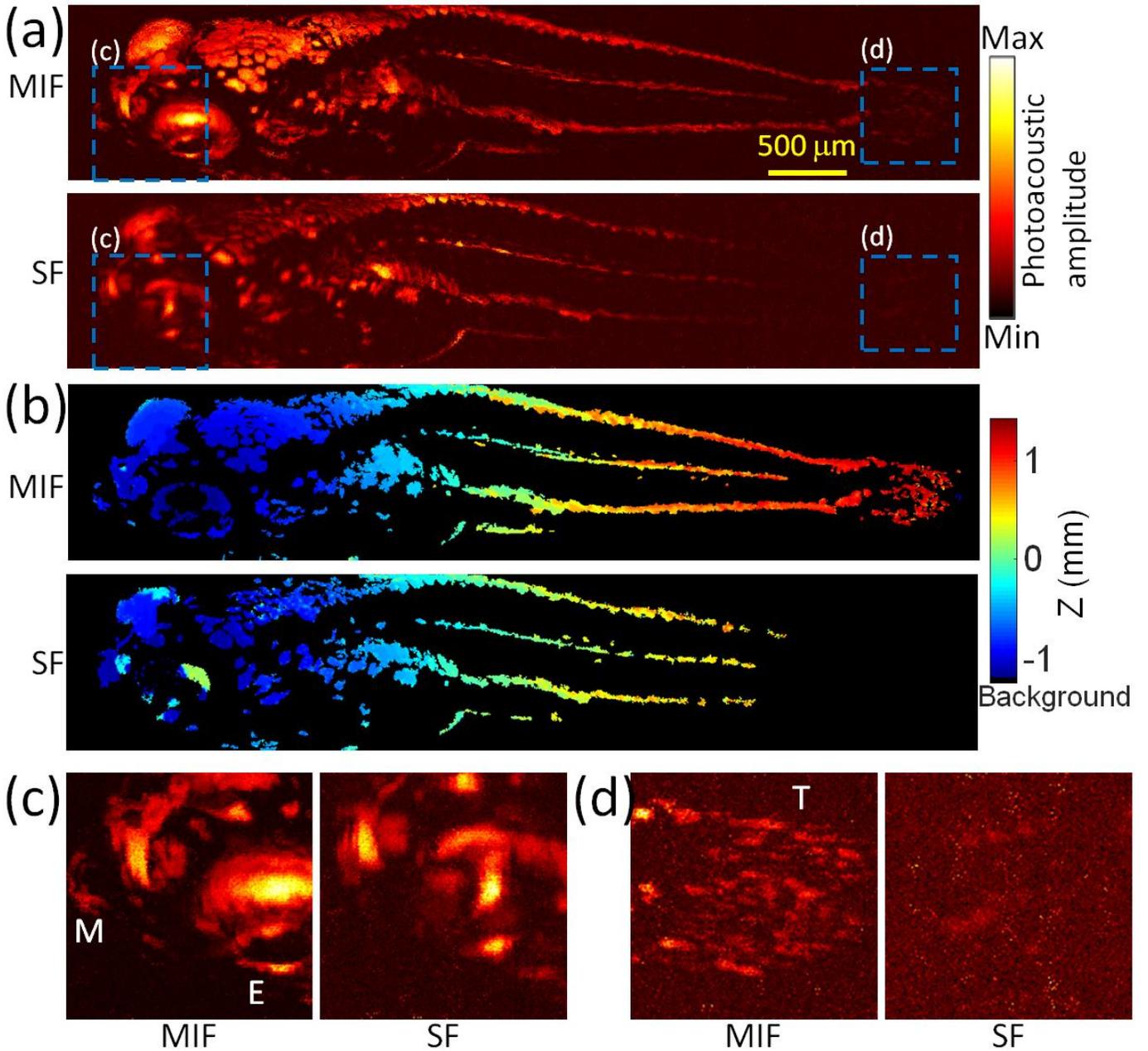

Fig. 10. (a) Photoacoustic XY MAP images of the MIF and SF images. (b) The depth-encoded images of (a). 3D videos of the MIF (Video 5) and SF (Video 6) images are available. Z represents the distance from the focal plane of the SF image ($f$ = 4.6 mm in this case). (c,d) Comparison of the zoom in images of the dashed boxes in (b). (c) for Z = −1.1 mm; (d) for Z = 1.1 mm. (a) and (b) share the same scale bar in (a). M, mouth; E, eye; T, tail.

## IV. DISCUSSION AND CONCLUSIONS

In conclusion, we developed a novel FA-PAE probe that achieved high resolution of 3−5 μm with ultra-large DOF of >3.2 mm, which was more than 27 times the SF-based DOF for SF-based lateral resolution of <5 μm. Imaging of the mouse eye, mouse ear, and zebrafish larvae was conducted to demonstrate *in vivo* imaging capability and excellent imaging performance of the probe. The unique optomechanical-based design for focus adjustment is immune against electromagnetic interference and the probe diameter is 2.9 mm, both facilitating clinical PAE applications. To our knowledge, our probe achieved unprecedented performance among existing OR-PAE probes when simultaneously taking resolution, DOF, and probe size into consideration. Although some efforts have been made to extend DOF for PAE recently [25], [27], the resolution was limited to >10 μm. Considering a focused Gaussian beam, DOF is proportional to the square of lateral resolution. Therefore, as resolution is enhanced (i.e., tighter focusing), the DOF drastically reduces. Existing methods suffer from the fundamental limitation (e.g., [27]), and are expected to largely sacrifice DOF when the resolution is further boosted to several micrometers. In contrast, our approach circumvents this limitation and thus, promising results can be achieved.

The MIF image requires image acquisition of several photoacoustic images at different focal planes to realize large-DOF imaging. For selected applications where only

SF-based scanning along the irregular surface of tissue (e.g., irregular gastrointestinal tract) is needed, one could consider photoacoustic signal feedback [23] or a water-balloon-based probe [26] for boundary recognition, which would greatly save the image acquisition time. It should be stressed that the above approach is intrinsically adaptive SF-based imaging without extended DOF, which is different from the MIF image with high resolution over large DOF by the proposed FA-PAE probe. In the above demonstrations, samples were basically with only slight scattering, such as the fish embedded in gel. It is expectable that the MIF image degrades in scattering media because of the strong scattering for deep focusing. To improve penetration depth, optical clearing could be considered for particular applications [28], [29]. Testing of the probe for endoscopic imaging of more tissues and animals would be of great interest for future work.

The 2-mm plano-convex lens was used in PAE for the first time. Another advantage of using the 2-mm plano-convex lens is low chromatic aberration compared with the common GRIN lens used in PAE, which facilitates functional imaging such as oxygen saturation ($sO_2$) measurement with high resolution and will be presented in detail in our another work in near future. Overall, our work made a step forward to PAE technologies in high microscale resolution over large DOF, which will greatly facilitate high-resolution PAE imaging applications such as angiogenesis studies of tumors. In addition to PAE, the novel optomechanical design for focus adjustment to obtain the MIF image may be exploited for other focusing-based endoscopic modalities, such as optical coherence tomography and confocal fluorescence microscopy, to enable high resolution over large DOF.